# L-PASSO: Adaptive Log-penalized Variable Selector, with Sparse Selection and Robustness to the Pilot Estimator

**Samiran Ghosh**[1], **Shuva Gupta**[2] **and Shreyo Ghosh**[3]

[1]*Department of Biostatistics and Data Science,*
*School of Public Health, The University of Texas Health Science Center at Houston*

[2]*University of Pennsylvania, Department of Statistics*

[3]*Department of Biomedical Engineering, Rutgers University–New Brunswick*

**Abstract:** We study L-PASSO, a penalized least-squares regression method based on the logarithmic penalty $\lambda \sum_j \log(1 + |\beta_j|/\tau)$, and its two-stage adaptive form, in which the scales $\tau_j$ come from a pilot estimator. Although this concave penalty has been proposed previously, its variable-selection properties and sparsity relative to other popular penalties have not been fully characterized. We derive the exact orthogonal-design solution, and show that the single-stage estimator selects the true model under a weighted irrepresentable condition that holds for every design once $\tau$ is small. We further establish that, under matched tuning, the two-stage estimator possesses the oracle property and has half the asymptotic bias of the adaptive LASSO. Our main theoretical contribution concerns the robustness of the two-stage procedure to the pilot estimator. We show that two-stage L-PASSO remains selection consistent even when the weights associated with relevant variables are as small as $\exp\{-o(1/\zeta_n)\}$, whereas the adaptive LASSO fails to retain relevant variables whose pilot weights are sufficiently small relative to its tuning parameter $\zeta_n$. We establish that both thresholds are sharp. In high-dimensional settings, we prove support recovery for a local minimizer without imposing any lower-bound condition on the weights of relevant variables, and provide conditions under which this local minimizer is also the unique global minimizer. We additionally establish the selection consistency of a BIC-based tuning-parameter selection procedure. Finally, simulations comparing L-PASSO with the LASSO, SCAD, MCP, SELO, uniLasso, and the adaptive LASSO show that L-PASSO most frequently recovers the true support. Two real-data examples further illustrate the method.

## 1. Introduction

The LASSO (Tibshirani, 1996) selects variables by penalizing $\sum_j |\beta_j|$. Its selection consistency requires the irrepresentable condition (Zhao and Yu, 2006), and when it selects correctly it shrinks every nonzero coefficient by the same amount, so it cannot be an oracle (Zou, 2006). The adaptive LASSO (Zou, 2006) repairs both of these defects with data-dependent weights $1/|\tilde{\beta}_j|^\gamma$ taken from a pilot estimator $\tilde{\beta}$; folded concave penalties such as SCAD (Fan and Li, 2001) and MCP (Zhang, 2010) repair them by flattening the penalty for large coefficients.

This paper studies the *least product absolute shrinkage and selection operator*, hence forward L-PASSO,

$$\widehat{\beta} \in \arg\min_{\beta \in \mathbb{R}^p} \left\{ \frac{1}{2n} \|y - X\beta\|^2 + \lambda \sum_{j=1}^{p} \log\left(1 + \frac{|\beta_j|}{\tau}\right) \right\}. \quad (1)$$

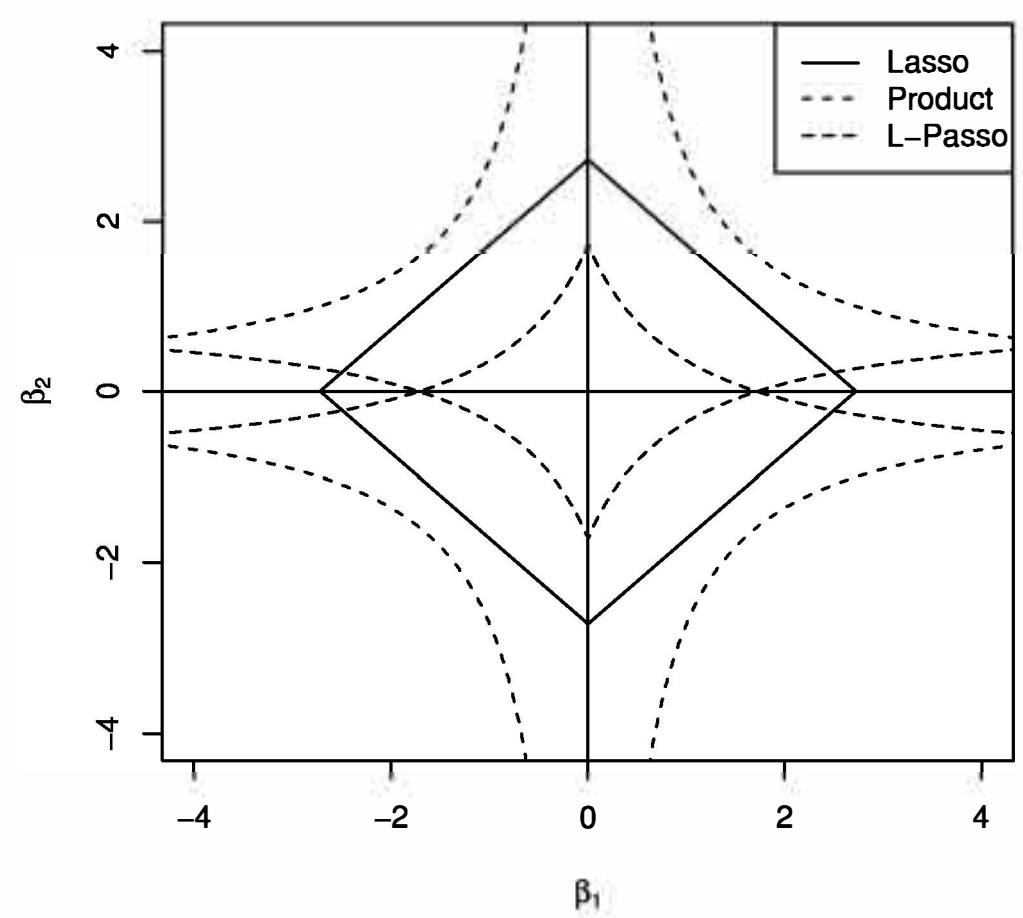

Fig 1. *Contours of the LASSO penalty, a simple product penalty, and the L-PASSO penalty in two dimensions.*

The name reflects that the penalty is the logarithm of the product $\prod_j (1 + |\beta_j|/\tau)$. Figure 1 compares its contours with those of the LASSO and of a simple product penalty. The underlying concave penalty is not new: it is the log-sum penalty used in reweighted $\ell_1$ minimization (Candès, Wakin and Boyd, 2008) and, up to reparameterization and an additive constant, corresponds to the negative log-density of the generalized double Pareto prior, whose maximum a posteriori estimator was studied by Armagan, Dunson and Lee (2013). However, the model-selection properties of this penalty, particularly in the two-stage adaptive setting, remain insufficiently understood. Local linear approximation (Zou and Li, 2008) turns it into a sequence of weighted LASSO problems. A single such step from the pilot is a weighted LASSO with weights $\zeta/(\tau_j + |\tilde{\beta}_j|)$, so, like the adaptive LASSO, it keeps a relevant input only if that input's weight is not too large; our tolerance results concern minimizers of the log-penalized objective itself, not its one-step approximation. What has been missing is a systematic theory of variable selection that disentangles the effects of the logarithmic penalty shape from those of adaptive weighting. This paper develops such a theory, characterizing the selection properties of both single-stage and two-stage L-PASSO and establishing when the logarithmic penalty offers advantages beyond those attributable to adaptive weighting alone.

**Contributions.**

1. *Structure (Section 2).* A difference-of-convex decomposition gives existence and a sharp uniqueness condition; an exact reparametrization turns the log penalty into an $\ell_1$ penalty; and the penalty bridges the LASSO ($\tau \to \infty$) and best-subset selection ($\tau \to 0$).
2. *Exact orthogonal-design solution (Section 3).* The thresholding rule has a convex regime, with a continuous rule and the LASSO's zero set at matched slope, and a nonconvex regime, with a jump, non-unique global minimizers, and first-order conditions that are not sufficient. In particular L-PASSO is *not* uniformly sparser than the LASSO; its robust advantage is smaller bias on large coefficients.

3. *Single stage (Section 4).* At the $\sqrt{n}$ scale, L-PASSO is not selection consistent. Beyond it, L-PASSO is selection consistent under a *weighted* irrepresentable condition, whose weights $\tau/(\tau + |\beta^*_j|)$ are below one; for every design it holds once $\tau$ is small relative to the signal.
4. *Two stages (Section 5).* With a $\sqrt{n}$-consistent pilot, two-stage L-PASSO is selection consistent and an oracle; at matched tuning its asymptotic bias is exactly half that of the adaptive LASSO. A BIC rule is established which chooses the tuning parameter consistently (Theorem 5.3).
5. *Pilot tolerance (Section 6), the main result.* Two-stage L-PASSO stays selection consistent when relevant pilot weights are as small as $\exp\{-o(1/\zeta_n)\}$; the adaptive LASSO drops any relevant input whose weight is small compared with $\zeta_n$. Both thresholds are sharp (Proposition 6.3). Screening-type pilots produce such weights naturally.
6. *High dimensions (Section 7).* A primal-dual witness argument gives support recovery for a strict local minimizer, with no condition on the relevant pilot weights, and an $\ell_0$-type argument gives conditions under which that minimizer is the unique global minimizer (Theorem 7.4).
7. *Comparisons and real data analysis (Sections 8.2 and 9).* Against the LASSO, SCAD, MCP, SELO, uniLasso and the adaptive LASSO in four designs, L-PASSO with a LASSO pilot and $\gamma = 2$ recovers the exact support most often and predicts as well as SCAD and MCP; the diabetes and prostate data exhibit the screening-pilot and sign-reversal problems.

**Relation to folded concave theory.** Fan, Xue and Zou (2014) showed that one step of local linear approximation from a LASSO start reaches the oracle estimator for folded concave penalties, whose derivative vanishes beyond a multiple of the tuning parameter. The log penalty does not satisfy this condition: its derivative never vanishes. Consequently, Fan's one-step oracle argument does not directly apply, and the selection properties of L-PASSO must be established through a direct analysis.

**Closely related work.** The seamless-$L_0$ penalty of Dicker, Huang and Lin (2013) is another smooth penalty that bridges $\ell_1$ and $\ell_0$, with consistent selection and asymptotic normality while $p$ grows more slowly than $n$; it is bounded and single-stage, so the questions about pilot weights studied here do not arise for it. Lv and Fan (2009) showed for a family of concave penalties that the LASSO's irrepresentable condition is relaxed by a factor determined by the penalty's derivatives at zero and at the signal; Theorem 4.4 is the exact fixed-dimension version for L-PASSO, together with a matching necessary condition. Zhang and Zhang (2012) gave a general theory under which global solutions of concave penalized least squares are oracle; it bounds the penalty's concavity by sparse eigenvalues, which fails for adaptive weights tending to zero, the regime of Theorem 7.4. Friedman (2012) developed path algorithms for log-type penalties.

**General Notation.** $y = X\beta^* + \varepsilon$ with $X \in \mathbb{R}^{n\times p}$, $S = \{j : \beta^*_j \neq 0\}$, $s = |S|$, $N = S^c$, $s_j = \mathrm{sign}(\beta^*_j)$, $\beta_{\min} = \min_{j\in S} |\beta^*_j|$, $\widehat{\Sigma} = X^\top X/n$, $L_n(\beta) = (2n)^{-1}\|y - X\beta\|^2$, and $\rho_\tau(t) = \log(1 + |t|/\tau)$ for $\tau > 0$, with $\rho_0(t) = +\infty \cdot 1\{t \neq 0\}$.

## 2. The estimator and its structure

**Definition 2.1** (L-PASSO)**.** The single-stage estimator is any global minimizer of $Q(\beta) = L_n(\beta) + \lambda \sum_j \rho_\tau(\beta_j)$. Given a pilot $\tilde{\beta}$ and $\gamma > 0$, the two-stage estimator is any global minimizer of $Q_2(\beta) = L_n(\beta) + \zeta \sum_j \rho_{\tau_j}(\beta_j)$ with $\tau_j = |\tilde{\beta}_j|^\gamma$. The adaptive LASSO with the same pilot minimizes $L_n(\beta) + \zeta \sum_j |\beta_j|/\tau_j$.

The two-stage penalty $\zeta \rho_{\tau_j}$ has slope $\zeta/\tau_j$ at zero, the adaptive LASSO weight, but slope $\zeta/(\tau_j + |\beta_j|)$ away from zero. This single difference drives Sections 5 and 6.

**Proposition 2.2** (Existence)**.** *For $\lambda > 0$ (respectively $\zeta > 0$) the set of global minimizers of $Q$ (respectively $Q_2$) is nonempty and compact.*

*Proof.* The objective is continuous on the closed set where it is finite and tends to $+\infty$ as $\|\beta\| \to \infty$, because $L_n \geq 0$ and each $\rho_\tau(\beta_j) \to \infty$ as $|\beta_j| \to \infty$. Its sublevel sets are therefore compact. □

**Proposition 2.3** (Difference of convex functions; uniqueness)**.** *For $\tau > 0$, $h_\tau(t) = \rho_\tau(t) + t^2/(2\tau^2)$ is convex, so*

$$Q(\beta) = \left[L_n(\beta) - \frac{\lambda}{2\tau^2}\|\beta\|^2\right] + \lambda \sum_j h_\tau(\beta_j).$$

*If $\lambda_{\min}(\widehat{\Sigma}) > \lambda/\tau^2$, then $Q$ is strictly convex, its global minimizer is unique, and the first-order conditions are necessary and sufficient for it.*

*Proof.* For $t > 0$, $h_\tau''(t) = -1/(\tau + t)^2 + 1/\tau^2 \geq 0$; at $t = 0$ the one-sided derivatives are $\mp 1/\tau$, a convex kink; and $h_\tau$ is even. The bracket has Hessian $\widehat{\Sigma} - (\lambda/\tau^2)I \succ 0$. □

When the condition fails, as it always does when $p > n$, global minimizers need not share fitted values or supports, and first-order conditions are not sufficient; Theorem 3.1(c) gives explicit examples even with a single predictor. Hence, properties of the LASSO that rest on convexity do not carry over.

**Proposition 2.4** (Exact $\ell_1$ reparametrization)**.** *Let $g(u) = \operatorname{sign}(u)(e^{|u|} - 1)$. Then $g$ is an odd, strictly increasing bijection of $\mathbb{R}$, continuously differentiable with $g'(u) = e^{|u|}$ and infinitely differentiable off 0, and its inverse is $g^{-1}(x) = \operatorname{sign}(x)\log(1 + |x|)$. With $\beta = \tau g(\nu)$,*

$$\rho_\tau(\beta_j) = |\nu_j|, \qquad Q(\tau g(\nu)) = L_n(\tau g(\nu)) + \lambda\|\nu\|_1.$$

*Hence $\widehat{\beta}$ minimizes $Q$ if and only if $\hat{\nu} = g^{-1}(\widehat{\beta}/\tau)$ minimizes $\nu \mapsto L_n(\tau g(\nu)) + \lambda\|\nu\|_1$, and $\widehat{\beta}$ and $\hat{\nu}$ have the same support and signs. Equivalently,*

$$\nu_j = \operatorname{sign}(\beta_j)\log\left(1 + \frac{|\beta_j|}{\tau}\right) = \log\frac{2\tau + |\beta_j| + \beta_j}{2\tau + |\beta_j| - \beta_j}.$$

*Proof.* The properties of $g$ are immediate. Since $|\tau g(\nu_j)|/\tau = e^{|\nu_j|} - 1$, $\rho_\tau(\tau g(\nu_j)) = |\nu_j|$. The map $\nu \mapsto \tau g(\nu)$ is a coordinatewise bijection that preserves zeros and signs, so minimizers correspond. For the last identity, if $\beta_j > 0$ the ratio is $(2\tau + 2\beta_j)/(2\tau) = 1 + \beta_j/\tau$, and if $\beta_j < 0$ it is the reciprocal of $1 + |\beta_j|/\tau$. □

The reparametrization exchanges a nonconvex penalty for a nonconvex loss, so it does not make the problem easier to optimize. Its value is as a device: zeros of $\beta$ and $\nu$ coincide, and the $\ell_1$ structure in $\nu$ makes the subdifferential at zero explicit.

**Proposition 2.5** (Bridge between the LASSO and best subset)**.** *(a) $\tau\rho_\tau(t) \to |t|$ as $\tau \to \infty$, uniformly on compacts; with $\lambda = \lambda'\tau$ the L-PASSO objective converges to the LASSO objective with parameter $\lambda'$. (b) $\rho_\tau(t)/\log(1/\tau) \to 1\{t \neq 0\}$ as $\tau \to 0$ for each $t$; with $\lambda = \lambda'/\log(1/\tau)$ the penalty converges pointwise to $\lambda'\|\beta\|_0$.*

*Proof.* (a) $\tau\log(1+|t|/\tau) = |t| - t^2/(2\tau) + O(\tau^{-2})$ uniformly for bounded $t$. (b) For $t \neq 0$, $\log(1+|t|/\tau)/\log(1/\tau) = 1 + \log(\tau+|t|)/\log(1/\tau) \to 1$, and the ratio is 0 at $t = 0$. □

*Remark* 2.6. The second-order term in (a) is $-t^2/(2\tau)$, a negative ridge term. The log penalty therefore does not approximate the ridge or the elastic net (Ghosh, 2011); it interpolates between $\ell_1$ and $\ell_0$.

**Proposition 2.7** (Optimality conditions)**.** *Let $\tau_j \geq 0$, let $A = \{j : \tau_j > 0\}$ (a weight $\tau_j = 0$ forces $\beta_j = 0$), and write $g_j(\beta) = X_j^\top(y - X\beta)/n$ and $a_j = \widehat{\Sigma}_{jj} > 0$.*

*(a)* First-order conditions. *Every local minimizer $\widehat{\beta}$ of $Q_2$ satisfies, for each $j \in A$,*

$$g_j(\widehat{\beta}) = \frac{\zeta\ \mathrm{sign}(\widehat{\beta}_j)}{\tau_j + |\widehat{\beta}_j|} \quad \text{if } \widehat{\beta}_j \neq 0, \qquad |g_j(\widehat{\beta})| \leq \frac{\zeta}{\tau_j} \quad \text{if } \widehat{\beta}_j = 0.$$

*(b)* Sufficiency under convexity. *If $\lambda_{\min}(\widehat{\Sigma}_{AA}) > \zeta/\min_{j\in A}\tau_j^2$, then $Q_2$ is strictly convex on $\{\beta : \beta_j = 0,\ j \notin A\}$, and the conditions in (a) are necessary and sufficient for its unique global minimizer.*

*(c)* A stronger necessary condition for global minimizers. *Every global minimizer is coordinatewise optimal: for each $j \in A$, $\widehat{\beta}_j$ minimizes $t \mapsto \frac{1}{2}(t - v_j)^2 + (\zeta/a_j)\rho_{\tau_j}(t)$ with $v_j = \widehat{\beta}_j + g_j(\widehat{\beta})/a_j$, so it is given by Theorem 3.1 with $v = v_j$ and $c = \zeta/a_j$.*

*(d)* Insufficiency in general. *Without the condition in (b), the conditions in (a) need not imply even (c): by Theorem 3.1(c), with a single predictor and $c > \tau^2$, $\widehat{\beta} = 0$ satisfies (a) whenever $|v| \leq c/\tau$, yet it is not a global minimizer when $|v| > v^*$.*

*Proof.* (a) If $\widehat{\beta}_j \neq 0$, $Q_2$ is differentiable in $\beta_j$ near $\widehat{\beta}_j$, with $\partial_j Q_2(\widehat{\beta}) = -g_j(\widehat{\beta}) + \zeta\,\mathrm{sign}(\widehat{\beta}_j)/(\tau_j + |\widehat{\beta}_j|)$, which must vanish. If $\widehat{\beta}_j = 0$, the function $t \mapsto Q_2(\widehat{\beta} + te_j)$ has right derivative $-g_j(\widehat{\beta}) + \zeta/\tau_j$ and left derivative $-g_j(\widehat{\beta}) - \zeta/\tau_j$ at 0, because $\rho_{\tau_j}$ has one-sided derivatives $\pm 1/\tau_j$ at zero; a local minimum requires the first to be nonnegative and the second nonpositive. (b) On $\{\beta_j = 0,\ j \notin A\}$, Proposition 2.3 applied coordinatewise gives $Q_2 = \left[L_n(\beta) - \frac{\zeta}{2}\sum_{j\in A}\beta_j^2/\tau_j^2\right] + \zeta\sum_{j\in A} h_{\tau_j}(\beta_j)$ with each $h_{\tau_j}$ convex, and the bracket has Hessian $\widehat{\Sigma}_{AA} - \zeta\,\mathrm{diag}(\tau_j^{-2}) \succeq \widehat{\Sigma}_{AA} - (\zeta/\min_A \tau_j^2)I \succ 0$. So $Q_2$ is strictly convex there, and (a) is exactly the condition that zero belongs to its subdifferential, which characterizes the unique minimizer. (c) A global minimizer also minimizes $Q_2$ along each coordinate line. Since $L_n(\widehat{\beta} + te_j) = L_n(\widehat{\beta}) - tg_j(\widehat{\beta}) + \frac{a_j}{2}t^2$, substituting $u = \widehat{\beta}_j + t$ and dividing by $a_j$ shows that $\widehat{\beta}_j$ minimizes $\frac{1}{2}(u - v_j)^2 + (\zeta/a_j)\rho_{\tau_j}(u)$. (d) In the scalar case $g = a(v - \widehat{\beta})$, so the condition at zero in (a) reads $|v| \leq c/\tau$; Theorem 3.1(b),(c) gives the rest. □

*Remark* 2.8. In the coordinates $\nu_j = \mathrm{sign}(\beta_j)\log(1+|\beta_j|/\tau_j)$ of Proposition 2.4, the conditions in (a) are exactly the subgradient conditions of the $\ell_1$ penalty $\zeta\|\nu\|_1$: since $\partial\beta_j/\partial\nu_j = \tau_j + |\beta_j|$, they read $g_j(\tau_j + |\beta_j|) = \zeta\,\mathrm{sign}(\nu_j)$ if $\nu_j \neq 0$ and $|g_j|\tau_j \leq \zeta$ if $\nu_j = 0$. The proofs below use these conditions as follows: Theorem 4.4 works in the convex case (b); the oracle expansion in Theorem 5.1 uses the equation in (a); the exclusion arguments of Theorems 5.1, 6.1 and 7.4 are coordinatewise comparisons in the spirit of (c); and Theorem 7.1 constructs a point satisfying (a) with strict inequality off the support. The coordinate descent algorithm of Section 3 returns points satisfying (c).

## 3. Exact solution under orthogonal design

If $\widehat{\Sigma} = I_p$, then $L_n(\beta) = \frac{1}{2}\|\beta - \tilde{\beta}\|^2 + \text{const}$ with $\tilde{\beta} = X^\top y/n$, and (1) separates into the scalar problems

$$\min_{t\in\mathbb{R}} G(t) = \tfrac{1}{2}(t-v)^2 + c\,\rho_\tau(t), \qquad v = \tilde{\beta}_j,\ c = \lambda. \tag{2}$$

For $v \geq 0$ (the case $v < 0$ follows by symmetry) put, when $(v+\tau)^2 \geq 4c$,

$$t_+(v) = \tfrac{1}{2}\Big[(v-\tau) + \sqrt{(v+\tau)^2 - 4c}\Big].$$

**Theorem 3.1** (Thresholding rule). *(a)* Convex regime, $c \leq \tau^2$. *The solution of* (2) *is unique and continuous in* $v$*: it is* $0$ *if* $|v| \leq c/\tau$ *and* $\mathrm{sign}(v)\, t_+(|v|)$ *otherwise.*

*(b)* Nonconvex regime, $c > \tau^2$. *There is a unique* $v^* \in (2\sqrt{c} - \tau,\ c/\tau)$ *such that the solution is* $0$ *for* $|v| < v^*$ *and* $\mathrm{sign}(v)\, t_+(|v|)$ *for* $|v| > v^*$*, and both are solutions at* $|v| = v^*$*. The rule jumps from* $0$ *to* $t_+(v^*) > 0$.

*(c)* Consequences in the nonconvex regime. *Global minimizers can be non-unique, with different supports and fitted values; and for* $v^* < |v| < c/\tau$, $t = 0$ *satisfies the first-order conditions without being a global minimizer.*

*(d)* Bias. *At a nonzero solution,* $|v| - |\hat{t}| = c/(\tau + |\hat{t}|)$*, which tends to* $0$ *as* $|\hat{t}| \to \infty$*; the LASSO's bias is the constant* $c$.

*Proof.* Take $v \geq 0$; a minimizer is nonnegative. For $t > 0$, $G'(t) = t - v + c/(\tau + t)$, and $G'(t) = 0$ if and only if $t^2 + (\tau - v)t + (c - v\tau) = 0$, whose discriminant is $(v+\tau)^2 - 4c$. When it is nonnegative, $t_+$ is the larger root and $G''(t_+) = 1 - c/(\tau + t_+)^2 \geq 0$ because $\tau + t_+ \geq (v+\tau)/2 \geq \sqrt{c}$, so $t_+$ is the only possible positive local minimizer.

(a) If $c \leq \tau^2$, then $G''(t) = 1 - c/(\tau+t)^2 > 0$ for $t > 0$, so $G$ is strictly convex on $[0,\infty)$. Zero is optimal if and only if the right derivative $-v + c/\tau$ is nonnegative. Otherwise $c - v\tau < 0$, so $t_+ > 0$; and $(v+\tau)^2 - 4c \geq (c/\tau - \tau)^2 \geq 0$ for $v \geq c/\tau$.

(b) If $v < 2\sqrt{c} - \tau$, the quadratic is positive, $G' > 0$ on $(0,\infty)$, and $0$ is the unique minimizer. At $v = 2\sqrt{c} - \tau$, $G' \geq 0$ with equality only at $t_+ = \sqrt{c} - \tau > 0$, so $G(t_+) > G(0)$, and by continuity this persists for slightly larger $v$. At $v = c/\tau$, $G'(0^+) = 0$ and $G''(0^+) = 1 - c/\tau^2 < 0$, so $0$ is not a local minimizer and $t_+$ wins. By the envelope theorem $\frac{d}{dv}\big[G(t_+(v)) - G(0)\big] = -(t_+ - v) - v = -t_+ < 0$, so the difference crosses zero exactly once, at some $v^* \in (2\sqrt{c} - \tau, c/\tau)$. Since $t_+ \geq \sqrt{c} - \tau > 0$ on this range, the rule jumps.

(c) At $|v| = v^*$, $0$ and $\mathrm{sign}(v)t_+$ are both global minimizers; in a design with one orthonormal column they give different fitted values. For $v^* < |v| < c/\tau$, $0$ satisfies $|v| \leq c/\tau$, the first-order condition at zero, but $G(t_+) < G(0)$.

(d) Rearrange $G'(\hat{t}) = 0$. □

**Corollary 3.2** (Comparison with the LASSO). *Under orthogonal design let the LASSO use penalty $\lambda' \sum_j |\beta_j|$, so it soft-thresholds at $\lambda'$.*

*(a)* Matched slope at zero, $\lambda' = c/\tau$. *In the convex regime the two estimators have the same zero set, and every nonzero L-PASSO estimate is larger in magnitude, with bias $c/(\tau + |\hat{t}|) < c/\tau$. In the nonconvex regime L-PASSO has weakly fewer zeros, since $v^* < c/\tau$.*

*(b)* Matched $\lambda$, $\lambda' = c$. *If $c \le \tau^2$ and $\tau \le 1$, L-PASSO is sparser, since $c/\tau \ge c$. Outside the convex regime the ordering can go either way: with $\tau = 0.9$, $c = 10$ and $v = 10$ the LASSO estimate is* 0 *while L-PASSO's is* 8.99*; with $\tau = 0.1$ and $c = 1$, $v^* \approx 2.52 > c$.*

*Proof.* (a) and the first claim of (b) follow from Theorem 3.1(a),(b),(d). The examples are computed from the closed form and confirmed by brute-force minimization. □

So "sparser than the LASSO" is not a theorem in general. What holds robustly is that, at the LASSO's zero set, L-PASSO's nonzero estimates are less biased.

**Computation.** For general designs we use coordinate descent with exact coordinate updates: with the other coordinates fixed, $Q$ is $\frac{a_j}{2}(\beta_j - v_j)^2 + \lambda\rho_\tau(\beta_j) + \text{const}$, where $a_j = \widehat{\Sigma}_{jj}$ and $v_j = \beta_j + X_j^\top r/(na_j)$ for the current residual $r$, and Theorem 3.1 with $c = \lambda/a_j$ gives its global minimizer in closed form. The objective decreases monotonically, and limit points are coordinatewise minima wherever the coordinate minimizers are unique (Tseng, 2001; Mazumder, Friedman and Hastie, 2011). Closed-form log thresholding also appears in iterative log-thresholding methods for sparse recovery (Malioutov and Aravkin, 2014).

## 4. The single-stage estimator in fixed dimension

**Assumption 4.1.** If $p$ and $\tau$ are fixed, $\widehat{\Sigma} \to C \succ 0$, $\max_i \|x_i\|^2/n \to 0$, and the $\varepsilon_i$ are i.i.d. with mean 0 and variance $\sigma^2$. Then $X^\top\varepsilon/\sqrt{n} \to W \sim N(0, \sigma^2 C)$ in distribution.

Write $w_j = s_j/(\tau + |\beta_j^*|)$ for $j \in S$, so that $\lambda w_j$ is the penalty's gradient at $\beta_j^*$.

**Theorem 4.2** ($\sqrt{n}$-scale limit). *Under Assumption 4.1, if $\sqrt{n}\lambda_n \to \lambda_0 \in [0, \infty)$, then every sequence of global minimizers satisfies $\sqrt{n}(\widehat{\beta} - \beta^*) \to \arg\min V$ in distribution, where*

$$V(u) = \tfrac{1}{2}u^\top Cu - u^\top W + \lambda_0\Big(\sum_{j\in S} w_j u_j + \frac{1}{\tau}\sum_{j\in N} |u_j|\Big).$$

*Proof.* We adapt the argument of Knight and Fu (2000) for the LASSO.

*Rate*. Let $\Delta = \widehat{\beta} - \beta^*$. From $Q(\widehat{\beta}) \le Q(\beta^*)$ and the fact that $\rho_\tau$ is $(1/\tau)$-Lipschitz, $\frac{1}{2}\Delta^\top\widehat{\Sigma}\Delta \le \Delta^\top X^\top\varepsilon/n + (\lambda_n/\tau)\|\Delta\|_1 \le \big(O_p(n^{-1/2}) + \lambda_n\sqrt{p}/\tau\big)\|\Delta\|$, so $\|\Delta\| = O_p(n^{-1/2})$.

*Convergence*. $\hat{u}_n = \sqrt{n}\Delta$ minimizes

$$\begin{aligned} V_n(u) &= n\big[Q(\beta^* + u/\sqrt{n}) - Q(\beta^*)\big] \\ &= \tfrac{1}{2}u^\top\widehat{\Sigma}u - u^\top X^\top\varepsilon/\sqrt{n} + n\lambda_n \textstyle\sum_j \big[\rho_\tau(\beta_j^* + u_j/\sqrt{n}) - \rho_\tau(\beta_j^*)\big]. \end{aligned}$$

For $j \in S$, a Taylor expansion gives $n\lambda_n[\cdots] = \sqrt{n}\lambda_n w_j u_j + O(\lambda_n u_j^2)$; for $j \in N$, $n\lambda_n \log(1 + |u_j|/(\sqrt{n}\tau)) = \sqrt{n}\lambda_n|u_j|/\tau + O(\lambda_n u_j^2)$. So $V_n$ converges to $V$ in distribution in $\ell^\infty(K)$ for every compact $K$, jointly through $X^\top\varepsilon/\sqrt{n} \to W$. $V$ is strictly convex, so its minimizer is unique, and $\hat{u}_n = O_p(1)$. The argmin continuous mapping theorem (van der Vaart and Wellner, 1996, Theorem 3.2.2) gives the claim. □

**Corollary 4.3** (No selection consistency at the $\sqrt{n}$ scale)**.** *Under the conditions of Theorem 4.2 with $N \neq \emptyset$, $\limsup_n \mathbb{P}(\widehat{S} = S) \leq q < 1$, and $q = 0$ if $\lambda_0 = 0$, where*

$$q = \mathbb{P}\Big(\big|Z + \lambda_0 C_{NS} C_{SS}^{-1} w_S\big| \leq \lambda_0/\tau \text{ componentwise}\Big), \qquad Z = W_N - C_{NS} C_{SS}^{-1} W_S.$$

*Proof.* Since $\{\widehat{S} = S\} \subseteq \{\hat{u}_{n,N} = 0\}$ and the set $F = \{u : u_N = 0\}$ is closed, the portmanteau theorem gives $\limsup \mathbb{P}(\widehat{S} = S) \leq \mathbb{P}(\arg\min V \in F)$. By the subgradient conditions, $\arg\min V$ has $u_N = 0$ exactly when $u_S = C_{SS}^{-1}(W_S - \lambda_0 w_S)$ and $|W_k - C_{kS} u_S| \leq \lambda_0/\tau$ for every $k \in N$, which is the event defining $q$. The Gaussian vector $Z$ has covariance $\sigma^2(C_{NN} - C_{NS} C_{SS}^{-1} C_{SN}) \succ 0$, so it lies in a bounded box with probability below one, and in a single point with probability zero. □

Beyond the $\sqrt{n}$ scale the picture changes. Define the *weighted irrepresentable coefficient*

$$\eta = \big\|C_{NS} C_{SS}^{-1} \omega_S\big\|_\infty, \qquad \omega_j = \tau w_j = \frac{\tau s_j}{\tau + |\beta_j^*|} \quad (j \in S).$$

The LASSO's coefficient (Zhao and Yu, 2006) has $s_j$ in place of $\omega_j$.

**Theorem 4.4** (Selection beyond the $\sqrt{n}$ scale)**.** *Under Assumption 4.1, let $\lambda_n \to 0$ and $\sqrt{n}\lambda_n \to \infty$.*

*(a) If $\eta < 1$, then $\mathbb{P}(\widehat{S} = S$ and $\operatorname{sign}(\widehat{\beta}_S) = s_S) \to 1$.*
*(b) If $\eta > 1$, then $\mathbb{P}(\widehat{S} = S) \to 0$.*
*(c) Under the condition of (a), $\widehat{\beta}_S - \beta_S^* = -\lambda_n C_{SS}^{-1} w_S + o_P(\lambda_n)$, so $\sqrt{n}(\widehat{\beta}_S - \beta_S^*)$ is not tight.*

*Proof.* Since $\lambda_n \to 0$ and $\lambda_{\min}(\widehat{\Sigma}) \to \lambda_{\min}(C) > 0$, Proposition 2.3 applies for $n$ large: $Q$ is strictly convex, its minimizer is unique, and the first-order conditions characterize it. Let $\check{b}$ minimize the restricted objective $L_n(b, 0) + \lambda_n \sum_{j \in S} \rho_\tau(b_j)$ over $b \in \mathbb{R}^s$, also strictly convex. As $\lambda_n \to 0$, $\check{b} \to \beta_S^*$ in probability, so with probability tending to one its signs are $s_S$ and its stationarity condition reads $\widehat{\Sigma}_{SS}(\check{b} - \beta_S^*) = X_S^\top \varepsilon/n - \lambda_n \check{w}_S$ with $\check{w}_j = s_j/(\tau + |\check{b}_j|) \to w_j$. This gives (c) on the event $\widehat{S} = S$, since $X_S^\top \varepsilon/n = O_P(n^{-1/2}) = o_P(\lambda_n)$. The vector $\check{\beta} = (\check{b}, 0)$ is the global minimizer if and only if $|X_k^\top(y - X_S\check{b})/n| \leq \lambda_n/\tau$ for every $k \in N$. Now

$$\frac{\tau}{\lambda_n} \cdot \frac{X_k^\top(y - X_S\check{b})}{n} = \frac{\tau}{\lambda_n}\Big(\frac{X_k^\top \varepsilon}{n} - \widehat{\Sigma}_{kS}\widehat{\Sigma}_{SS}^{-1}\frac{X_S^\top \varepsilon}{n}\Big) + \widehat{\Sigma}_{kS}\widehat{\Sigma}_{SS}^{-1}(\tau \check{w}_S) \to \big(C_{NS} C_{SS}^{-1} \omega_S\big)_k$$

in probability, because the first term is $O_P(1/(\sqrt{n}\lambda_n)) = o_P(1)$. (a) If $\eta < 1$, the inequalities hold strictly with probability tending to one, so $\check{\beta}$ is the minimizer. (b) If the $k$-th limit exceeds one in absolute value, then on $\{\widehat{S} = S\}$ the minimizer must equal $\check{\beta}$, whose condition at $k$ fails with probability tending to one. □

*Remark* 4.5 (Where exactly the log shape helps). Each $|\omega_j| < 1$, and $\eta \leq \|C_{NS} C_{SS}^{-1}\|_\infty \tau/(\tau + \beta_{\min})$. Hence for every design with $C \succ 0$ there is $\tau_0 > 0$ such that $\eta < 1$ for all $\tau < \tau_0$: single-stage L-PASSO can recover the support in designs where the LASSO provably cannot, for any tuning. Part (c) shows the price: the bias is of order $\lambda_n$, so no single-stage tuning gives both selection consistency and $\sqrt{n}$-consistency (Corollary 4.3). At finite $n$, small $\tau$ also places the problem in the nonconvex regime of Section 3.

## 5. The two-stage estimator

**Theorem 5.1** (Selection consistency and oracle property)**.** *Under Assumption 4.1, let the pilot satisfy $\sqrt{n}(\tilde\beta - \beta^*) = O_p(1)$, let $\gamma > 0$, and let $\zeta_n \to 0$ with $\sqrt{n}\zeta_n \to \kappa \in [0,\infty)$ and $\zeta_n n^{(1+\min\{\gamma,1\})/2} \to \infty$. Then every sequence of global minimizers of $Q_2$ satisfies $\mathbb{P}(\widehat{S} = S) \to 1$ and*

$$\sqrt{n}(\widehat\beta_S - \beta^*_S) \to N\big(-\kappa C_{SS}^{-1} w^\circ,\ \sigma^2 C_{SS}^{-1}\big), \qquad w^\circ_j = \frac{s_j}{|\beta^*_j|^\gamma + |\beta^*_j|}.$$

*With $\kappa = 0$ this is the oracle property.*

*Proof.* *Rate.* For $j \in S$, $\tau_j \to |\beta^*_j|^\gamma > 0$, so $\rho_{\tau_j}$ is $O_p(1)$-Lipschitz. From $Q_2(\widehat\beta) \le Q_2(\beta^*)$, dropping the nonnegative penalty on $N$, $\frac{1}{2}\Delta^\top\widehat\Sigma\Delta \le \Delta^\top X^\top\varepsilon/n + \zeta_n O_p(1)\|\Delta\|$, so $\|\Delta\| = O_p(n^{-1/2} + \zeta_n) = O_p(n^{-1/2})$.

*Exclusion.* Fix $k \in N$ with $\widehat\beta_k \neq 0$ and compare $\widehat\beta$ with $\widehat\beta - \widehat\beta_k e_k$. Since

$$L_n(\beta - he_k) = L_n(\beta) + hX_k^\top(y - X\beta)/n + \tfrac{h^2}{2}\widehat\Sigma_{kk},$$

optimality gives $\zeta_n\rho_{\tau_k}(\widehat\beta_k) \le |\widehat\beta_k|\big(|X_k^\top(y - X\widehat\beta)/n| + \frac{1}{2}\widehat\Sigma_{kk}|\widehat\beta_k|\big)$. Using $\rho_\tau(t) \ge |t|/(\tau + |t|)$ and dividing by $|\widehat\beta_k|$,

$$\frac{\zeta_n}{\tau_k + |\widehat\beta_k|} \le \left|\frac{X_k^\top\varepsilon}{n} - \widehat\Sigma_{k\cdot}\Delta\right| + \widehat\Sigma_{kk}|\widehat\beta_k| = O_p(n^{-1/2}). \tag{3}$$

But $\tau_k + |\widehat\beta_k| = O_p(n^{-\min\{\gamma,1\}/2})$, so the left side is at least of order $\zeta_n n^{\min\{\gamma,1\}/2}$, which is $\gg n^{-1/2}$. Hence $\mathbb{P}(\widehat\beta_N = 0) \to 1$; and $\widehat\beta_S \neq 0$ with probability tending to one by consistency.

*Expansion.* On $\{\widehat S = S\}$, $\widehat\beta_S$ is an interior stationary point: $\widehat\Sigma_{SS}(\widehat\beta_S - \beta^*_S) = X_S^\top\varepsilon/n - \zeta_n\tilde w_S$ with $\tilde w_j = s_j/(\tau_j + |\widehat\beta_j|) \to w^\circ_j$. Multiply by $\sqrt{n}$ and apply Slutsky's lemma. □

**Proposition 5.2** (Half the bias)**.** *Under the conditions of Theorem 5.1, the adaptive LASSO with the same pilot and $\zeta_n$ is also selection consistent, with limit $N(-\kappa C_{SS}^{-1}w^{\ell_1}, \sigma^2 C_{SS}^{-1})$, where $w^{\ell_1}_j = s_j/|\beta^*_j|^\gamma$. For $\gamma = 1$, $w^\circ = \frac{1}{2}w^{\ell_1}$ exactly. The two penalties have the same slope $\zeta_n/\tau_j$ at zero, hence the same exclusion condition; so at matched tuning L-PASSO has half the asymptotic bias, and to first order it coincides with the adaptive LASSO run at $\zeta_n/2$.*

*Proof.* The proof of Theorem 5.1 applies with $\zeta_n|t|/\tau_j$ in place of $\zeta_n\rho_{\tau_j}(t)$: the rate step is unchanged, the exclusion step is (3) with $\tau_k$ in place of $\tau_k + |\widehat\beta_k|$, and the stationarity condition has $\zeta_n s_j/\tau_j$. For $\gamma = 1$, $|\beta^*_j| + |\beta^*_j| = 2|\beta^*_j|$. □

In simulation (Section 8) the ratio of mean biases was 0.51 to 0.52 against the theoretical 0.5.

**Theorem 5.3** (BIC tuning)**.** *Under Assumption 4.1, let $\widehat\beta_\zeta$ be a global minimizer of $Q_2$ at tuning value $\zeta$, $\widehat S_\zeta$ its support, $\hat\sigma^2_\zeta = \|y - X\widehat\beta_\zeta\|^2/n$, and $\mathrm{BIC}(\zeta) = \log\hat\sigma^2_\zeta + |\widehat S_\zeta|\log n/n$. Let $\mathcal{Z}_n$ be any set of tuning values that contains a sequence $\zeta^*_n$ satisfying the conditions of Theorem 5.1, and let $\hat\zeta$ minimize BIC over $\mathcal{Z}_n$. Then $\mathbb{P}(\widehat S_{\hat\zeta} = S) \to 1$.*

*Proof.* We follow Wang, Li and Tsai (2007). Write $\mathrm{RSS}_T$ for the least-squares residual sum of squares on a support $T$. By Theorem 5.1, $\mathbb{P}(\widehat{S}_{\zeta_n^*} = S) \to 1$; on that event the stationarity condition gives $\widehat{\beta}_{\zeta_n^*,S} - \hat{\beta}_S^{\mathrm{OLS}} = O_p(\zeta_n^*)$, and since least-squares residuals on $S$ are orthogonal to $X_S$, $\hat{\sigma}^2_{\zeta_n^*} = \mathrm{RSS}_S/n + O_p(\zeta_n^{*2}) = \mathrm{RSS}_S/n + O_p(1/n)$. Hence $\mathrm{BIC}(\zeta_n^*) = \log\sigma^2 + o_p(1)$. Every other support is either *underfitted* ($S \nsubseteq T$) or *overfitted* ($S \subsetneq T$), and the bounds below depend only on the support, so they hold uniformly over $\zeta \in \mathcal{Z}_n$. If $S \nsubseteq \widehat{S}_\zeta$, then $\hat{\sigma}^2_\zeta \geq \min_{T:S\nsubseteq T} \mathrm{RSS}_T/n \to \sigma^2 + c$ with $c = \min_{T:S\nsubseteq T} \min_{b:\mathrm{supp}(b)\subseteq T} (\beta^* - b)^\top C(\beta^* - b) > 0$, since there are finitely many such $T$ and $C \succ 0$; so $\mathrm{BIC}(\zeta) > \mathrm{BIC}(\zeta_n^*)$ with probability tending to one. If $S \subsetneq \widehat{S}_\zeta$, then $\hat{\sigma}^2_\zeta \geq \mathrm{RSS}_{\{1,\ldots,p\}}/n$, and $\mathrm{RSS}_S - \mathrm{RSS}_{\{1,\ldots,p\}} = O_p(1)$, a quadratic form in $\varepsilon$ of rank $p - s$. So $\log\hat{\sigma}^2_{\zeta_n^*} - \log\hat{\sigma}^2_\zeta \leq O_p(1/n)$, while $|\widehat{S}_\zeta| - s \geq 1$; hence $\mathrm{BIC}(\zeta) - \mathrm{BIC}(\zeta_n^*) \geq \log n/n - O_p(1/n) > 0$ with probability tending to one. Since $\mathrm{BIC}(\hat{\zeta}) \leq \mathrm{BIC}(\zeta_n^*)$, $\widehat{S}_{\hat{\zeta}} = S$ with probability tending to one. □

The same argument applies to the adaptive LASSO. For weights as in Theorem 6.1, it applies whenever $\mathcal{Z}_n$ contains a $\zeta_n^*$ satisfying that theorem with $\zeta_n^* = o(\sqrt{\log n/n})$, so that the fit on $S$ costs $o_p(\log n/n)$ in $\hat{\sigma}^2$. In an AR(0.5) design with $p = 10$, $\beta^* = (2, -1.5, 1, 0, \ldots, 0)$ and an OLS pilot, BIC-tuned L-PASSO selected exactly $S$ in 89%, 92% and 97% of 100 data sets at $n = 100$, 400 and 1600, and the BIC-tuned adaptive LASSO in 84%, 94% and 97%.

## 6. Tolerance of small pilot weights

Theorem 5.1 assumes a $\sqrt{n}$-consistent pilot, so every relevant weight $\tau_j$ stays bounded away from zero. Pilots in practice need not do this: a screening pilot built from marginal regressions, a heavily penalized LASSO, or any pilot facing weak signals can return a tiny weight for a relevant input. Write $\underline{\tau}_n = \min_{j\in S} \tau_j$ and $\bar{\tau}_n = \max_{k\in N} \tau_k$, for weights from any pilot.

**Theorem 6.1** (L-PASSO tolerates small relevant weights)**.** *Under Assumption 4.1, let the weights $\tau_j > 0$ be arbitrary random variables with, in probability,*

*(i)* $\zeta_n \to 0$ *and* $\zeta_n \log(1/\underline{\tau}_n) \to 0$*;*
*(ii)* $\bar{\tau}_n \to 0$*;*
*(iii)* $\sqrt{n}\zeta_n/(\bar{\tau}_n + n^{-1/2}) \to \infty$.

*Then every sequence of global minimizers of $Q_2$ satisfies $\mathbb{P}(\widehat{S} = S) \to 1$ and $\|\widehat{\beta} - \beta^*\| = O_p(n^{-1/2} + \zeta_n)$. No lower bound on $\underline{\tau}_n$ beyond (i) is needed.*

*Proof. Consistency.* Since

$$P_S(\beta^*) = \sum_{j\in S} \rho_{\tau_j}(\beta_j^*) \leq s\log\big(1 + \max_j |\beta_j^*|/\underline{\tau}_n\big),$$

optimality against $\beta^*$ gives $\frac{1}{2}\Delta^\top\widehat{\Sigma}\Delta \leq \Delta^\top X^\top\varepsilon/n + \zeta_n P_S(\beta^*)$, and the right side is $O_p(n^{-1/2})\|\Delta\| + o_p(1)$ by (i). Hence $\Delta \to 0$ and $|\widehat{\beta}_j| \geq \beta_{\min}/2$ for $j \in S$ with probability tending to one.

*Rate, free of the relevant weights.* For $x, y > 0$, $\log(x/y) \leq |x - y|/\min\{x, y\}$. With $x = \tau_j + |\beta_j^*|$ and $y = \tau_j + |\widehat{\beta}_j|$, this gives $|\rho_{\tau_j}(\beta_j^*) - \rho_{\tau_j}(\widehat{\beta}_j)| \leq 2|\Delta_j|/\beta_{\min}$ for $j \in S$, whatever $\tau_j$ is. So $\frac{1}{2}\lambda_{\min}(\widehat{\Sigma})\|\Delta\|^2 \leq \big(O_p(n^{-1/2}) + 2\sqrt{s}\zeta_n/\beta_{\min}\big)\|\Delta\|$ and $\|\Delta\| = O_p(n^{-1/2} + \zeta_n)$.

*Exclusion.* Inequality (3) holds for any weights, and its right side is now at most $C_1(n^{-1/2}+\zeta_n)$ with probability tending to one, as is $|\widehat{\beta}_k|$. If $\widehat{\beta}_k \neq 0$ for some $k \in N$, then $\zeta_n \leq \big(\bar{\tau}_n + C_1(n^{-1/2} + \zeta_n)\big) C_1(n^{-1/2} + \zeta_n)$, that is,

$$1 \leq C_1\big(\bar{\tau}_n + C_1 n^{-1/2} + C_1\zeta_n\big)\Big(1 + \frac{n^{-1/2}}{\zeta_n}\Big).$$

The right side tends to zero: the first factor does by (i) and (ii), and the cross term $(\bar{\tau}_n + n^{-1/2})\, n^{-1/2}/\zeta_n$ does by (iii). This is a contradiction, so $\widehat{\beta}_N = 0$ with probability tending to one. □

**Proposition 6.2** (The adaptive LASSO does not tolerate them)**.** *Under Assumption 4.1, let $\widehat{\beta}$ minimize $L_n(\beta) + \zeta_n \sum_j |\beta_j|/\tau_j$. If $\zeta_n/\tau_j \to \infty$ in probability for some $j \in S$, then $\mathbb{P}(\widehat{\beta}_j = 0) \to 1$. Consequently $\mathbb{P}(\widehat{S} = S) \to 0$ and $\|\widehat{\beta} - \beta^*\| \geq |\beta^*_j|$ with probability tending to one.*

*Proof.* Optimality against 0 gives $L_n(\widehat{\beta}) \leq L_n(0) = O_p(1)$, hence $\|X\widehat{\beta}\|/\sqrt{n} = O_p(1)$ and $\|\widehat{\beta}\| = O_p(1)$ because $\lambda_{\min}(\widehat{\Sigma}) \to \lambda_{\min}(C) > 0$. On a ball containing $\widehat{\beta}$, $|X_j^{\top}(y - X\beta)/n| + \widehat{\Sigma}_{jj}\|\beta\| \leq C_L$ with $C_L = O_p(1)$. Moving $\widehat{\beta}_j$ to zero raises $L_n$ by at most $C_L|\widehat{\beta}_j|$ and lowers the penalty by $(\zeta_n/\tau_j)|\widehat{\beta}_j|$, which is larger with probability tending to one unless $\widehat{\beta}_j = 0$. □

**Proposition 6.3** (Sharpness)**.** *Under Assumption 4.1, if $\zeta_n$ is bounded and $\zeta_n \log(1/\tau_j) \to \infty$ in probability for some $j \in S$, then every sequence of global minimizers of $Q_2$ has $\mathbb{P}(\widehat{\beta}_j = 0) \to 1$.*

*Proof.* As in the proof of Proposition 6.2, $\|\widehat{\beta}\| = O_p(1)$, and moving $\widehat{\beta}_j$ to zero raises $L_n$ by at most $C_L|\widehat{\beta}_j|$ with $C_L = O_p(1)$, while it lowers the penalty by $\zeta_n \rho_{\tau_j}(\widehat{\beta}_j)$. Since $\zeta_n \log(1/\tau_j) \to \infty$ and $\zeta_n$ is bounded, $\tau_j \to 0$. Suppose $\widehat{\beta}_j \neq 0$. If $|\widehat{\beta}_j| \geq \sqrt{\tau_j}$, the saving is at least $\zeta_n \log(1 + \tau_j^{-1/2}) \geq \frac{1}{2}\zeta_n \log(1/\tau_j)$, which exceeds the bounded increase $C_L|\widehat{\beta}_j|$. If $|\widehat{\beta}_j| < \sqrt{\tau_j}$, the saving is at least $\zeta_n|\widehat{\beta}_j|/(\tau_j + |\widehat{\beta}_j|) \geq \zeta_n|\widehat{\beta}_j|/(2\sqrt{\tau_j})$, and $\zeta_n/\sqrt{\tau_j} = \zeta_n \log(1/\tau_j)/(\sqrt{\tau_j}\log(1/\tau_j)) \to \infty$ because $\sqrt{t}\log(1/t) \leq 2/e$; so it exceeds $C_L|\widehat{\beta}_j|$. Either way $\widehat{\beta}$ is not a global minimizer, with probability tending to one. □

*Remark* 6.4 (A sharp dichotomy). Theorem 6.1 and Proposition 6.3 show that L-PASSO keeps a relevant input when $\zeta_n \log(1/\tau_j) \to 0$ and drops it when $\zeta_n \log(1/\tau_j) \to \infty$. For the adaptive LASSO the threshold is $\zeta_n/\tau_j$: Proposition 6.2 drops the input when $\zeta_n/\tau_j \to \infty$, and when $\zeta_n/\tau_j \to 0$ for all $j \in S$ the consistency step of Theorem 6.1, with penalty $\zeta_n \sum_{j\in S} |\beta^*_j|/\tau_j \to 0$, keeps every relevant input. The two thresholds differ exponentially.

**Corollary 6.5** (Exponential gap)**.** *Suppose $\bar{\tau}_n = O_p(n^{-1/2})$ and $\underline{\tau}_n \geq n^{-a}$ for a fixed $a > 0$. Two-stage L-PASSO is selection consistent for every $\zeta_n$ with $n\zeta_n \to \infty$ and $\zeta_n \log n \to 0$. The adaptive LASSO drops every relevant input whose weight is of order $n^{-a}$ as soon as $n^a\zeta_n \to \infty$. In general, L-PASSO tolerates relevant weights down to $\exp\{-o(1/\zeta_n)\}$, and the adaptive LASSO only down to order $\zeta_n$.*

*Proof.* Conditions (i) and (ii) of Theorem 6.1 read $\zeta_n a \log n \to 0$ and $\bar{\tau}_n \to 0$. Condition (iii) holds because $\bar{\tau}_n + n^{-1/2} = O_p(n^{-1/2})$, so $\sqrt{n}\zeta_n/(\bar{\tau}_n + n^{-1/2})$ is of order $n\zeta_n \to \infty$. The adaptive LASSO claim is Proposition 6.2. □

**Example 6.6** (Screening pilots)**.** Let the pilot be the vector of marginal regression coefficients $\tilde\beta_j = X_j^\top y/\|X_j\|^2$, with $\gamma = 1$, as in screening-guided methods. Suppose $\widehat\Sigma - C = O_p(n^{-1/2})$, as for random designs, the irrelevant inputs are uncorrelated with the relevant ones ($C_{NS} = 0$), and some relevant input $j_0$ is marginally uncorrelated with the response, $(C\beta^*)_{j_0} = 0$, although $\beta^*_{j_0} \neq 0$. Then $\tau_k = O_p(n^{-1/2})$ for $k \in N$ and $\tau_{j_0} = O_p(n^{-1/2})$ with $\log(1/\tau_{j_0}) = O_p(\log n)$. By Corollary 6.5, L-PASSO is selection consistent whenever $n\zeta_n \to \infty$ and $\zeta_n \log n \to 0$, while the adaptive LASSO drops $j_0$ whenever $\sqrt{n}\zeta_n \to \infty$. Its sufficient conditions for keeping $j_0$ and for excluding $N$ leave only the window $n^{-1} \ll \zeta_n \ll n^{-1/2}$.

## 7. High dimensions

Let $p$ be arbitrary, with columns normalized so that $\|X_j\|^2/n \le 1$. Put $\kappa = \lambda_{\min}(\widehat\Sigma_{SS})$, $\kappa' = \kappa - 4\zeta/\beta^2_{\min}$, and the incoherence $\vartheta = \max_{k\in N}\|\widehat\Sigma_{kS}\|_2$, which is at most $\sqrt{s}$ and is typically much smaller. For constants $\mu, \mu_S > 0$ consider the event

$$\mathcal{E} = \left\{\|X_N^\top \varepsilon/n\|_\infty \le \mu,\ \ \|X_S^\top \varepsilon/n\|_2 \le \mu_S\right\}.$$

**Theorem 7.1** (Primal-dual witness)**.** *Let all weights $\tau_j > 0$, let $r = 2(\mu_S + \zeta\sqrt{s}/\beta_{\min})/\kappa'$, and suppose*

$$\text{(C1) } r < \beta_{\min}/2, \qquad \text{(C2) } \kappa' > 0, \qquad \text{(C3) } \zeta/\bar\tau > \mu + \vartheta r.$$

*On $\mathcal{E}$, $Q_2$ has a strict local minimizer $\check\beta$ with $\operatorname{supp}(\check\beta) = S$, $\operatorname{sign}(\check\beta_S) = s_S$, and $\|\check\beta - \beta^*\|_2 \le r$. No condition is placed on $\tau_j$ for $j \in S$.*

*Proof. Restricted problem.* On $B = \{b \in \mathbb{R}^s : \|b - \beta^*_S\|_\infty \le \beta_{\min}/2\}$ every $|b_j| \ge \beta_{\min}/2$, so $\rho''_{\tau_j}(b_j) \ge -1/(\tau_j + |b_j|)^2 \ge -4/\beta^2_{\min}$, and $F(b) = L_n(b,0) + \zeta\sum_{j\in S}\rho_{\tau_j}(b_j)$ is $\kappa'$-strongly convex on $B$. Let $\check b$ minimize $F$ over $B$. Strong convexity and $F(\check b) \le F(\beta^*_S)$ give $\|\check b - \beta^*_S\| \le 2\|\nabla F(\beta^*_S)\|/\kappa'$, and $\nabla F(\beta^*_S) = -X_S^\top\varepsilon/n + \zeta\left(s_j/(\tau_j + |\beta^*_j|)\right)_j$ has norm at most $\mu_S + \zeta\sqrt{s}/\beta_{\min}$. So $\|\check b - \beta^*_S\| \le r < \beta_{\min}/2$ by (C1): $\check b$ is interior, $\nabla F(\check b) = 0$, and its signs are $s_S$.

*Strict dual feasibility.* Let $\check\beta = (\check b, 0)$. For $k \in N$, $|\partial_k L_n(\check\beta)| \le |X_k^\top\varepsilon/n| + \|\widehat\Sigma_{kS}\|_2\|\check b - \beta^*_S\|_2 \le \mu + \vartheta r$. By (C3), $\delta_0 = \min_{k\in N}(\zeta/\tau_k - |\partial_k L_n(\check\beta)|) > 0$.

*Strict local minimality.* For small $\Delta$ with $\check b + \Delta_S \in B$, expand the quadratic $L_n$ in the $N$-block. Since $\frac{1}{2}\Delta_N^\top\widehat\Sigma_{NN}\Delta_N \ge 0$,

$$Q_2(\check\beta+\Delta) - Q_2(\check\beta) \ge [F(\check b + \Delta_S) - F(\check b)] + \sum_{k\in N}\left[\zeta\rho_{\tau_k}(\Delta_k) - |\partial_k L_n(\check\beta)||\Delta_k|\right] - |\Delta_N^\top\widehat\Sigma_{NS}\Delta_S|.$$

The first bracket is at least $\frac{\kappa'}{2}\|\Delta_S\|^2$. Using $\rho_\tau(t) \ge |t|/\tau - t^2/(2\tau^2)$, each term of the sum is at least $\frac{\delta_0}{2}|\Delta_k|$ once $|\Delta_k| \le \delta_0\tau_k^2/\zeta$. The last term is at most $\vartheta\|\Delta_N\|_1\|\Delta_S\| \le \frac{\delta_0}{4}\|\Delta_N\|_1$ once $\|\Delta_S\| \le \delta_0/(4\vartheta)$. So the difference is at least $\frac{\kappa'}{2}\|\Delta_S\|^2 + \frac{\delta_0}{4}\|\Delta_N\|_1 > 0$ for small $\Delta \neq 0$. □

**Corollary 7.2.** *With Gaussian noise and*

$$\mu = \sigma\sqrt{\frac{2\log(4p/\delta)}{n}}, \qquad \mu_S = \sigma\left(\sqrt{\frac{s}{n}} + \sqrt{\frac{2\lambda_{\max}(\widehat\Sigma_{SS})\log(2/\delta)}{n}}\right),$$

*$\mathbb{P}(\mathcal{E}) \ge 1 - \delta$. So under (C1) to (C3) the conclusion of Theorem 7.1 holds with probability at least $1 - \delta$.*

*Proof.* Each $X_k^\top \varepsilon/n$ is normal with variance at most $\sigma^2/n$; a Gaussian tail bound and a union bound over at most $p$ coordinates give $\mathbb{P}(\|X_N^\top \varepsilon/n\|_\infty > \mu) \le \delta/2$. The map $z \mapsto \|\sigma X_S^\top z/n\|_2$ of a standard normal vector $z$ is Lipschitz with constant $\sigma\sqrt{\lambda_{\max}(\widehat{\Sigma}_{SS})/n}$ and has mean at most $\sigma\sqrt{\mathrm{tr}(\widehat{\Sigma}_{SS})/n} \le \sigma\sqrt{s/n}$, so Gaussian concentration gives $\mathbb{P}(\|X_S^\top \varepsilon/n\|_2 > \mu_S) \le \delta/2$. □

*Remark* 7.3. *Comparison.* For the adaptive LASSO the same construction has stationarity term $\zeta s_j/\tau_j$ on $S$ and a convex objective, so the radius becomes $2(\mu_S + \zeta\|(1/\tau_j)_{j\in S}\|_2)/\kappa$, and (C1) then requires every relevant weight to be large compared with $\zeta/(\kappa\beta_{\min})$. Theorem 7.1 needs nothing of the kind.

*Numerical check.* In 20 simulated data sets ($n = 400$, $p = 1000$, $s = 5$, nonzero coefficients of size 1.5 to 2, $\sigma = 1$, LASSO pilot, $\zeta = 0.05$, and in half of them one relevant weight set to $10^{-6}$), conditions (C1) to (C3) held in all 20 (at this penalty the LASSO pilot selected no irrelevant input, so (C3) held with room to spare), every random perturbation of $\check{\beta}$ increased $Q_2$, and $\check{\beta}$ had support exactly $S$. With the worst-case bounds $\vartheta \le \sqrt{s}$ and $\mu_S \le \sqrt{s}\mu$ in place of the sharper quantities, the conditions held in none of the 20, so the sharper form matters.

*Computation.* In all 20 data sets above, proximal gradient started at the LASSO fit converged to $\check{\beta}$. Theorem 7.4 below gives conditions under which $\check{\beta}$ is the unique global minimizer.

### 7.1. *The global minimizer*

Global theory for concave penalties (Zhang and Zhang, 2012) bounds the penalty's concavity by a sparse eigenvalue. With adaptive weights the concavity of $\zeta\rho_{\tau_k}$ is $\zeta/\tau_k^2$, unbounded as $\tau_k \to 0$. We use the opposite feature: a small weight makes every nonzero irrelevant coefficient expensive, much as an $\ell_0$ penalty does.

Assume Gaussian noise and $\|X_j\|^2/n \le 1$. A weight $\tau_k = 0$ forces $\beta_k = 0$; let $N^+ = \{k \in N : \tau_k > 0\}$, $\bar{\tau} = \max_{k\in N^+}\tau_k$ ($\bar{\tau} = 0$ if $N^+$ is empty), $\underline{\tau} = \min_{j\in S}\tau_j > 0$, $B = \|\beta^*\|_\infty$ and $\underline{L} = \log(1 + B/\underline{\tau})$. For $\delta \in (0,1)$ put $x = \log(2/\delta)$, $\bar{\sigma}^2 = \sigma^2(1 + 2\sqrt{x/n} + 2x/n)$,

$$A_0 = \frac{\sigma^2(2s + 3\log(4/\delta))}{n}, \qquad A_1 = \frac{\sigma^2(5\log(ep) + 6)}{n}, \qquad g = \sqrt{\bar{\sigma}^2 + 2\zeta s\underline{L}} + \tfrac{1}{2},$$

$$\Lambda_N = \log\Big(1 + \frac{\zeta}{2g\bar{\tau}}\Big), \qquad \bar{m} = \frac{2(A_0 + \zeta s\underline{L})}{\zeta\Lambda_N}, \qquad R = 2\Big(\frac{A_0 + \bar{m}A_1 + \zeta s\underline{L}}{\phi}\Big)^{1/2},$$

with $\Lambda_N = \infty$ and $\bar{m} = 0$ when $\bar{\tau} = 0$, and $\phi = \min\{\lambda_{\min}(\widehat{\Sigma}_{TT}) : S \subseteq T \subseteq S \cup N^+,\ |T| \le s + \bar{m}\}$.

**Theorem 7.4** (Global minimizer). *Suppose*

*(G0)* $\bar{\tau} \le \zeta/(2g)$ *and* $\zeta \le 2g$;
*(G1)* $\zeta\Lambda_N \ge 2A_1$;
*(G2)* $R < \beta_{\min}/2$;
*(G3)* $\zeta\Lambda_N > 2A_0 + 16\zeta^2 s/(\phi\beta_{\min}^2)$;
*(G4)* $\kappa' = \lambda_{\min}(\widehat{\Sigma}_{SS}) - 4\zeta/\beta_{\min}^2 > 0$.

*With probability at least* $1-\delta$, $Q_2$ *has a unique global minimizer. It has support* $S$ *and signs* $s_S$, *satisfies* $\|\widehat{\beta} - \beta^*\|_2 \le R$, *and equals* $(\check{b}, 0)$, *where* $\check{b}$ *minimizes* $F$ *over* $B = \{b : \|b - \beta_S^*\|_\infty \le \beta_{\min}/2\}$ *as in the proof of Theorem 7.1. No lower bound on* $\underline{\tau}$ *is required beyond its logarithmic entry through* $\underline{L}$.

*Proof.* *Events.* For $M \subseteq N$ with $|M| = m$, let $P_{S\cup M}$ project onto the span of $X_{S\cup M}$ and $\psi_m = \max_{|M|=m} \|P_{S\cup M}\varepsilon\|^2/n$. By the chi-squared tail bound of Laurent and Massart (2000), for fixed $M$ and $x_m = m\log(ep) + \log(4(m+1)^2/\delta)$, $\|P_{S\cup M}\varepsilon\|^2 \le \sigma^2(2(s+m)+3x_m)$ except with probability $e^{-x_m}$. A union bound over the at most $(ep)^m$ sets of size $m$, summed over $m \ge 0$, fails with probability at most $\sum_m \delta/(4(m+1)^2) < \delta/2$. On the complement, using $\log(ep) \ge 1$ and $\log(m+1) \le m$, $\psi_m \le A_0 + mA_1$ for every $m$. The same tail bound gives $\|\varepsilon\|^2/n \le \bar{\sigma}^2$ except with probability $\delta/2$. We work on both events.

*Step 1: each nonzero irrelevant coefficient costs at least $\zeta\Lambda_N$.* Let $\widehat{\beta}$ be a global minimizer and $k \in N$ with $\widehat{\beta}_k \ne 0$, so $k \in N^+$. Comparing $\widehat{\beta}$ with $\widehat{\beta} - \widehat{\beta}_k e_k$, as in (3), gives $\zeta/(\tau_k + |\widehat{\beta}_k|) \le |\partial_k L_n(\widehat{\beta})| + \frac{1}{2}|\widehat{\beta}_k|$, and $|\partial_k L_n(\widehat{\beta})| \le \|y - X\widehat{\beta}\|/\sqrt{n} \le \sqrt{2Q_2(\widehat{\beta})} \le \sqrt{2Q_2(\beta^*)} \le g - \frac{1}{2}$. If $|\widehat{\beta}_k| \ge 1$, then $\rho_{\tau_k}(\widehat{\beta}_k) \ge \log(1 + 1/\bar{\tau}) \ge \Lambda_N$, using $\zeta \le 2g$. Otherwise $\zeta/(\tau_k + |\widehat{\beta}_k|) \le g$, so $|\widehat{\beta}_k| \ge \zeta/g - \bar{\tau} \ge \zeta/(2g)$ by (G0), and $\rho_{\tau_k}(\widehat{\beta}_k) \ge \Lambda_N$.

*Step 2: counting.* Let $M$ be the irrelevant support of $\widehat{\beta}$, $m = |M|$, $T = S \cup M$ and $\Delta = \widehat{\beta} - \beta^*$, supported on $T$. From $Q_2(\widehat{\beta}) \le Q_2(\beta^*)$ and $\langle \varepsilon, X\Delta\rangle/n = \langle P_T\varepsilon, X\Delta\rangle/n \le \|P_T\varepsilon\|^2/n + \|X\Delta\|^2/(4n)$,

$$\tfrac{1}{4}\|X\Delta\|^2/n + \zeta P_N(\widehat{\beta}) \le \psi_m + \zeta\left[P_S(\beta^*) - P_S(\widehat{\beta})\right]. \tag{4}$$

With Step 1, $P_S(\widehat{\beta}) \ge 0$ and $P_S(\beta^*) \le s\underline{L}$, this gives $\zeta m\Lambda_N \le A_0 + mA_1 + \zeta s\underline{L}$, and (G1) yields $m \le \bar{m}$.

*Step 3: localization.* Since $|T| \le s + \bar{m}$ and $T \subseteq S \cup N^+$, $\|X\Delta\|^2/n \ge \phi\|\Delta\|^2$, and (4) gives $\frac{1}{4}\phi\|\Delta\|^2 \le A_0 + \bar{m}A_1 + \zeta s\underline{L}$, that is $\|\Delta\| \le R < \beta_{\min}/2$ by (G2). Hence $|\widehat{\beta}_j| > \beta_{\min}/2$ for $j \in S$, with sign $s_j$.

*Step 4: no irrelevant coefficient.* Now $\log(x/y) \le |x - y|/\min\{x, y\}$ gives $P_S(\beta^*) - P_S(\widehat{\beta}) \le (2/\beta_{\min})\|\Delta_S\|_1 \le (2\sqrt{s}/\beta_{\min})\|\Delta\|$, so (4) becomes $\zeta m\Lambda_N + \frac{1}{4}\phi\|\Delta\|^2 \le A_0 + mA_1 + (2\zeta\sqrt{s}/\beta_{\min})\|\Delta\| \le A_0 + mA_1 + \frac{1}{8}\phi\|\Delta\|^2 + 8\zeta^2 s/(\phi\beta_{\min}^2)$. With (G1), $\frac{1}{2}\zeta\Lambda_N m \le A_0 + 8\zeta^2 s/(\phi\beta_{\min}^2)$; if $m \ge 1$ this contradicts (G3). So $m = 0$.

*Step 5: identification.* Thus $\widehat{\beta} = (\widehat{\beta}_S, 0)$ with $\widehat{\beta}_S \in B$, and $\widehat{\beta}_S$ minimizes $F$ over $\mathbb{R}^s$, hence over $B$. By (G4), $F$ is strictly convex on $B$, so $\widehat{\beta}_S = \check{b}$. Global minimizers exist (Proposition 2.2), and each equals $(\check{b}, 0)$. □

*Remark* 7.5. *Role of the weights.* Condition (G0) asks the pilot's false positives to have weights small compared with $\zeta$, and (G2) asks $\zeta s\underline{L}$ to be small compared with $\phi\beta_{\min}^2$, the usual signal-strength requirement; the relevant weights enter only through $\underline{L} = \log(1 + B/\underline{\tau})$, the global form of Section 6. Powers $\gamma > 1$ shrink false-positive weights quadratically, which helps (G0). Unlike Theorem 7.1, no dual condition such as (C3) is needed.

*Numerical check.* In eight data sets with $n = 5000$, $p = 2000$, $s = 5$, nonzero coefficients of size 1.5 to 2, $\sigma = 1$, $\delta = 0.05$, and a LASSO pilot at penalty $\sigma\sqrt{\log p/n}$, conditions (G0) to (G4) held for some $\zeta$ in all eight when $\gamma = 2$, with 6 to 17 false positives in the pilot and $R$ between 0.52 and 0.68, against $\beta_{\min}/2 = 0.75$. With $\gamma = 1$ they held in none: false positives then have weights of order 0.01, too large for (G0) at any $\zeta$ allowed by (G2). In 66 runs of proximal gradient from eleven starting points in each of six data sets with $n = 400$ and $p = 1000$, where the conditions do not hold, no run reached a lower objective than $\check{\beta}$. The probability bound $\psi_m \le A_0 + mA_1$ of the proof was checked directly by enumerating all subsets with $m \le 3$ for $n = 60$, $p = 16$, $s = 3$: in 2000 noise draws it was never violated, and the largest ratio of $\psi_m$ to its bound was 0.96.

## 8. Simulations

### 8.1. *Checks of the theory*

All estimators in this subsection were computed by coordinate descent with exact coordinate updates (Section 3). Two-stage fits used $\gamma = 1$ and were started at the least-squares fit, so that each method is compared at the same well-fitting starting point.

**A. Weighted irrepresentable condition (Theorem 4.4).** $S = \{1,2\}, \beta^* = (1,1,0)$, independent standard normal $Z_1, Z_2, Z_3$, $X_j = Z_j$ for $j = 1,2$, and $X_3 = 0.7Z_1 + 0.7Z_2 + \sqrt{0.52}\, Z_3$, with $\sigma = 1$. The LASSO's irrepresentable coefficient is $1.40 > 1$; L-PASSO's weighted coefficient at $\tau = 0.5$ is $1.40 \cdot 0.5/1.5 = 0.47 < 1$. Both used $\lambda_n = 0.5n^{-1/3}$, matched in slope at zero. Over 100 replications:

| $n$ | 200 | 1000 | 5000 |
|---|---|---|---|
| LASSO, $\mathbb{P}(\widehat{S} = S)$ | 0.33 | 0.22 | 0.10 |
| L-PASSO, $\mathbb{P}(\widehat{S} = S)$ | 0.84 | 0.90 | 0.91 |

**B. Pilot tolerance with a screening pilot (Theorem 6.1, Example 6.6).** $p = 20$, $\beta^* = (1,1,1,1,1,0,\dots,0)$, independent standard normal $Z_1,\dots,Z_{20}$, $X_j = Z_j$ except $X_5 = -(Z_1 + \cdots + Z_4)/4 + (\sqrt{3}/2)Z_5$, which has unit variance and zero covariance with the response although $\beta_5^* = 1$. The pilot is the vector of marginal regression coefficients. Over 50 replications, the counts are (runs keeping $X_5$) / (runs recovering exactly $S$):

| $n$ | 500 | 2000 | 8000 |
|---|---|---|---|
| median pilot weight on $X_5$ | 0.063 | 0.025 | 0.014 |
| $\zeta_n = 0.5\, n^{-1/2} \log n$ *(L-PASSO's window only)* | | | |
| adaptive LASSO | 0 / 0 | 0 / 0 | 0 / 0 |
| L-PASSO | 23 / 23 | 40 / 40 | 50 / 50 |
| $\zeta_n = 0.5\, n^{-3/4}$ *(inside both windows)* | | | |
| adaptive LASSO | 46 / 4 | 50 / 9 | 50 / 16 |
| L-PASSO | 50 / 3 | 50 / 11 | 50 / 16 |

At the small tuning value both methods keep $X_5$ but exclude the irrelevant inputs slowly, since $n\zeta_n$ grows only like $n^{1/4}$. Reliable recovery requires the larger tuning value, which only L-PASSO can use.

**C. Half the bias (Proposition 5.2).** $n = 2000, p = 8, \beta^* = (2,-1.5,1,0,\dots,0)$, independent standard normal inputs, $\sigma = 1$, least-squares pilot, $\zeta_n = 2/\sqrt{n}$, 200 replications. The mean biases on the three nonzero coefficients were $(-0.0229, 0.0309, -0.0457)$ for the adaptive LASSO and $(-0.0117, 0.0160, -0.0235)$ for L-PASSO; the ratios were 0.51, 0.52 and 0.51, against the theoretical 0.5. In absolute terms, Theorem 5.1 and Proposition 5.2 predict biases $-\kappa n^{-1/2} w^{\circ}$ and $-\kappa n^{-1/2} w^{\ell_1}$ with $\kappa = 2$, that is $(-0.0112, 0.0149, -0.0224)$ and $(-0.0224, 0.0298, -0.0447)$; the simulated values agree to within 8%.

**D. The $\sqrt{n}$-scale bound (Corollary 4.3).** In the design of A with $\tau = 0.5$ and $\lambda_n = 0.3/\sqrt{n}$, the bound is $q = 0.560$. Over 400 replications the frequency of exact selection was 0.547 at $n = 2000$ and 0.530 at $n = 20000$, consistent with $\limsup \mathbb{P}(\widehat{S} = S) \le q$.

**E. Sharpness (Proposition 6.3).** With $p = 5$, $\beta^* = (1,1,1,0,0)$, independent standard normal inputs, $n = 2000$, $\zeta = 0.1$ and the weight of input 1 set to $\exp(-c/\zeta)$, the fraction of 40 data sets in which input 1 was kept by the global minimizer (found by multi-start) was 1.00, 1.00, 0.00 and 0.00 for L-PASSO at $c = \zeta \log(1/\tau_1) = 0.05, 0.3, 1$ and 3, and 1.00, 0.00, 0.00 and 0.00 for the adaptive LASSO. L-PASSO switches once $\zeta \log(1/\tau_1)$ exceeds the input's contribution to the fit; the adaptive LASSO switches far earlier.

### 8.2. *Comparison with competing methods*

We compared ten estimators: the LASSO; SCAD ($a = 3.7$) and MCP ($\gamma = 3$), computed along the penalty path with warm starts; SELO (Dicker, Huang and Lin, 2013) with $\tau_0 = 0.01$; uniLasso, implemented as in Chatterjee, Hastie and Tibshirani (2025) (leave-one-out univariate fits, then a nonnegative LASSO on those fits, so that univariate signs are preserved); the adaptive LASSO with a LASSO pilot and with a marginal (screening) pilot, $\gamma = 1$; and two-stage L-PASSO with a LASSO pilot ($\gamma = 1$ and $\gamma = 2$) and with a marginal pilot ($\gamma = 1$). The SELO and L-PASSO fits were started at the LASSO fit and computed by proximal gradient with exact univariate steps, so that either can move a coefficient to zero in one step. Every method was tuned on an independent validation sample of the same size; Dicker, Huang and Lin (2013) recommend a BIC rule for SELO, which may favor it in selection. Prediction error is $(\widehat{\beta} - \beta^*)^\top \Sigma (\widehat{\beta} - \beta^*)$, computed from the population covariance. Each design has $n = 200$, $p = 300$, $s = 5$, $\sigma = 1$, and 30 replications with fixed random seeds: (1) independent inputs, $\beta^*_S = (2, -2, 1.5, -1.5, 1)$; (2) AR(1) inputs with correlation 0.5 and relevant inputs at positions $1, 3, 5, 7, 9$; (3) cancellation: $X_5 = -(X_1 + \cdots + X_4)/4 + (\sqrt{3}/2)Z_5$, $\beta^*_S = (1, \ldots, 1)$, so $X_5$ is marginally uncorrelated with the response; (4) sign reversal: $\mathrm{corr}(X_1, X_2) = 0.8$ with $\beta^*_1 = 2$, $\beta^*_2 = -1$, so $X_2$'s marginal effect is positive, and $\beta^*_{3:5} = (1.5, -1.5, 1)$.

| | independent | AR(0.5) | cancellation | sign reversal |
|---|---|---|---|---|
| LASSO | 0 / .143 | 0 / .218 | 0 / .256 | 0 / .305 |
| SCAD | 77 / .026 | 50 / .031 | 37 / .031 | 30 / .031 |
| MCP | 87 / .026 | 67 / .030 | 70 / .031 | 50 / .031 |
| SELO | 93 / .027 | 50 / .072 | 0 / .150 | 57 / .057 |
| uniLasso | 47 / .036 | 0 / .069 | 0 / .689 | 0 / .415 |
| adaptive LASSO, LASSO pilot | 60 / .026 | 33 / .033 | 20 / .044 | 23 / .055 |
| adaptive LASSO, marginal pilot | 10 / .044 | 0 / .094 | 0 / .450 | 0 / .128 |
| L-PASSO, LASSO pilot, $\gamma = 1$ | 90 / .023 | 73 / .032 | 87 / .036 | 70 / .041 |
| L-PASSO, LASSO pilot, $\gamma = 2$ | **97** / .023 | **83** / .030 | **93** / .033 | **97** / .033 |
| L-PASSO, marginal pilot | 43 / .028 | 20 / .048 | 10 / .060 | 17 / .064 |

Each cell gives the percentage of replications recovering exactly $S$, and the mean prediction error. L-PASSO with a LASSO pilot and $\gamma = 2$ recovered the exact support most often in every design, had the fewest false positives (at most 0.25 on average), and predicted as well as SCAD and MCP, within 0.003. With $\gamma = 1$ it recovered the support more often than every competing method in three designs and was second to SELO with independent inputs, with slightly larger prediction error in the correlated designs. SELO was strong with independent inputs but recovered the support in none of the cancellation replications. With the screening pilot, L-PASSO was far better than the adaptive LASSO, which averaged 32 false positives in the cancellation design against 2.6. The uniLasso estimator failed where marginal and partial effects disagree: it found all signs correctly in 23% of replications under cancellation and in none under sign reversal.

## 9. Data examples

### *9.1. The diabetes data*

The diabetes data of Efron et al. (2004) record a measure of disease progression one year after baseline for $n = 442$ patients, with ten baseline predictors: age, sex, body mass index (bmi), mean arterial pressure (bp), and six serum measurements s1 to s6. We standardized the predictors and centered the response.

**The pilot problem in real data.** The data contain exactly the situation of Example 6.6. Least-squares (partial) and marginal regression coefficients differ sharply for several inputs, and s1 and s2 are correlated at 0.90:

| | age | sex | s1 | s2 | s3 |
|---|---|---|---|---|---|
| least-squares coefficient | –0.5 | −11.4 | –37.7 | 22.7 | 4.8 |
| marginal coefficient | 14.5 | 3.3 | 16.3 | 13.4 | –30.4 |

A screening pilot therefore gives sex, which has a sizeable partial effect, a weight of 3.3 against 45.2 for bmi, and gives s1 and s3 the wrong sign. A method that preserves univariate signs, such as uniLasso, cannot let s1 enter with its partial sign.

**Methods.** We compared the LASSO with the adaptive LASSO and two-stage L-PASSO, each with a least-squares (OLS) pilot and with a marginal (screening) pilot, all with $\gamma = 1$. Penalties were chosen by five-fold cross-validation over a common grid of 22 values, geometrically spaced from 200 to 0.05,, and two-stage fits were started at the least-squares fit.

**Selection on the full data.** The cross-validated fits selected:

| method | selected inputs (sign) | size |
|---|---|---|
| LASSO | sex−, bmi, bp, s1−, s3−, s4, s5, s6 | 8 |
| adaptive LASSO, OLS pilot | sex−, bmi, bp, s1−, s2+, s4, s5, s6 | 8 |
| adaptive LASSO, marginal pilot | all ten, including age | 10 |
| L-PASSO, OLS pilot | sex−, bmi, bp, s1−, s2+, s4, s5, s6 | 8 |
| L-PASSO, marginal pilot | sex−, bmi, bp, s1−, s3−, s4, s5, s6 | 8 |

With the marginal pilot, cross-validation chose $\zeta = 3.85$ for the adaptive LASSO, which keeps all ten inputs, and $\zeta = 12.6$ for L-PASSO, which keeps sex and drops age. At a common $\zeta$ the two methods nearly agree: at $\zeta = 12.6$ they select the same eight inputs, and at $\zeta = 3.85$ they differ only in s3, which L-PASSO drops. So the tolerance mechanism of Section 6 does not operate in these data, because sex's weight is not small enough relative to $\zeta$. The difference comes through tuning: at a given $\zeta$, L-PASSO shrinks retained coefficients half as much (Proposition 5.2), which is consistent with cross-validation favoring a larger penalty for it, and the larger penalty gives sparser models.

**Stability.** Over 200 bootstrap resamples, each with its own marginal pilot and each method at its cross-validated $\zeta$, L-PASSO kept sex with a negative sign in 100% of resamples against 92% for the adaptive LASSO, kept s1 with a negative sign in 95% against 98%, included age in 72% against 91%, and selected 8.3 inputs on average against 9.1.

**Prediction.** Over random 75/25 splits, with tuning by five-fold cross-validation inside each training set, mean test error and model size were:

| method | 10 predictors (12 splits) test MSE (se) | size | 64 predictors (10 splits) test MSE (se) | size |
|---|---|---|---|---|
| LASSO | 3015 (98) | 8.4 | 3000 (93) | 18.8 |
| adaptive LASSO, OLS pilot | 3012 (101) | 7.6 | 3226 (103) | 28.3 |
| adaptive LASSO, marginal pilot | 3028 (100) | 8.5 | 3120 (91) | 18.6 |
| L-PASSO, OLS pilot | 3014 (100) | 7.3 | 3226 (101) | 26.3 |
| L-PASSO, marginal pilot | 3022 (101) | 8.2 | 3271 (124) | 27.8 |

The 64-predictor design adds the 45 pairwise interactions and the squares of the nine non-binary predictors, as in Efron et al. (2004). With ten predictors all methods predict equally well, and L-PASSO with the OLS pilot does so with the smallest models. With 64 correlated predictors the LASSO predicts best: pilots estimated from 332 training observations are noisy, and the two-stage estimators inherit that noise. Our theory concerns selection, and this example shows its limits for prediction.

### 9.2. *The prostate cancer data*

The prostate data of Stamey et al. (1989), used by Tibshirani (1996) to introduce the LASSO, relate log prostate-specific antigen to eight clinical measures for $n = 97$ men. Three inputs reverse sign between their partial (least-squares) and marginal coefficients, on the standardized scale: age (−0.15 partial, +0.19 marginal), lcp, the log capsular penetration (−0.15, +0.63), and gleason (+0.03, +0.42). We fitted the eight estimators of Section 8.2, with an OLS pilot in place of the LASSO pilot, tuning by five-fold cross-validation.

The uniLasso fit selected lcavol, lweight, lbph, svi and pgg45 and excluded lcp and age; its sign constraint cannot represent their negative partial effects. The other methods kept lcp and age with negative signs. Over 60 bootstrap resamples, each re-tuned, lcp entered with its partial sign in 0% of uniLasso fits, 40% of LASSO fits, and 55% to 68% of the fits of the remaining methods (55% and 63% for L-PASSO with the OLS and marginal pilots). Over 30 random splits into 73 training and 24 test observations, test error ranged from 0.588 (uniLasso) to 0.637 (MCP), with standard errors near 0.025; L-PASSO's errors were 0.624 and 0.606. No method predicted significantly better, and L-PASSO showed no advantage over the adaptive LASSO in these data. The example illustrates the cost of sign constraints for interpretation rather than for prediction.

## 10. Discussion

**Relation to other estimators.** The Dantzig selector (Candès and Tao, 2007) satisfies oracle inequalities of the same form as the LASSO's (Bickel, Ritov and Tsybakov, 2009), and its selection behavior parallels the LASSO's, including failures under correlated designs (Meinshausen, Rocha and Yu, 2007); Theorem 4.4 shows that L-PASSO relaxes exactly this kind of condition. SCAD and MCP flatten the penalty and so remove bias entirely for large coefficients; the log penalty keeps a bias of order $\zeta_n/(\tau_j + |\beta_j|)$ but charges only logarithmically for including an input, which is what produces the tolerance in Section 6. In the comparison of Section 8.2 the remaining bias cost little: with $\gamma = 2$, L-PASSO predicted as well as SCAD and MCP while recovering the exact support most often in every design. The uniLasso method (Chatterjee, Hastie and Tibshirani, 2025) fits univariate regressions first and preserves their signs in a second-stage fit; like any method guided by marginal fits, it is exposed to inputs that are marginally uncorrelated with the response (Example 6.6), and its sign constraint cannot accommodate an input whose partial effect reverses its marginal one, as s1 does in the diabetes data. In our simulations it never recovered all signs under sign reversal, and in the prostate data it excluded lcp and age. An interesting direction for future research is to develop a single penalty function that combines MCP's exact unbiasedness for large coefficients with L-PASSO's super-sparsity and robustness for small coefficients. The performance of L-PASSO in combination with screening methods, (e.g., SIS, DC-SIS, ExSIS, and ISIS) has not been investigated in this paper and remains an important direction for future research.

**Limitations.** The global high-dimensional result has conservative constants and, in our checks, applies with $\gamma = 2$ weights but not with $\gamma = 1$. With $\gamma = 1$, L-PASSO's remaining bias costs a little prediction accuracy against SCAD and MCP in correlated designs (Gupta, 2012); $\gamma = 2$ removed most of this gap in our simulations. In the diabetes data the differences between the two-stage estimators are modest and arise through tuning rather than through the tolerance mechanism; they did not improve prediction, and with 64 predictors they predicted worse than the LASSO. The tuning conditions are asymptotic, and at finite $n$ small $\tau$ or large $\zeta$ puts the problem in the nonconvex regime, where Theorem 3.1 shows that the solution can jump. Theorem 5.3 justifies BIC tuning in fixed dimension only; high-dimensional tuning, extensions to generalized linear models, and inference after selection are left for future work.

## Disclosure of AI use

The authors used Claude, a large language model developed by Anthropic and accessed through claude.ai in 2026, used to check internal consistency, revise, and organize the paper, code and to improve clarity of writing. The authors have verified the manuscript, reference against its original source, reviewed and reran all code. The authors take full responsibility for the content of this paper.

## Code and data availability

The diabetes data are distributed with scikit-learn, and the prostate data with the Rdatasets archive. Python code reproducing every simulation, numerical check and data analysis in this paper, with fixed random seeds, is provided as supplementary material and will be deposited in a public repository.